\documentclass[american,aps,pra,reprint,superscriptaddress,tikz]{revtex4-1}

\usepackage{tikz}
\usepackage{amsmath,amscd,amsthm,amssymb}
\usepackage{mathrsfs}
\usepackage{graphicx}
\usepackage{epsf,epstopdf}
\usepackage[center]{caption2}
\usepackage[all]{xy}
\usepackage[unicode=true,pdfusetitle, bookmarks=true,bookmarksnumbered=false,bookmarksopen=false, breaklinks=false,pdfborder={0 0 0},backref=false,colorlinks=false] {hyperref}
\hypersetup{colorlinks,linkcolor=myurlcolor,citecolor=myurlcolor,urlcolor=myurlcolor}
\usepackage{graphics,epstopdf,graphicx, amsthm, amsmath, amssymb, times, braket, color, bm}
\usepackage[up]{subfigure}
\usepackage{cleveref}

\usepackage{makecell}

\definecolor{myurlcolor}{rgb}{0,0,0.7}

\theoremstyle{plain}
\newtheorem{thm}{\protect\theoremname}

\newtheorem{lem}[thm]{Lemma}
\providecommand{\theoremname}{Theorem}
\newcommand*{\myproofname}{Proof}

\makeatother

\theoremstyle{definition}

\theoremstyle{remark}

\begin{document}

\author{Chunhe Xiong}
\email{xiongchunhe@zju.edu.cn}
\affiliation{School of Computer and Computing Science, Zhejiang University City College, Hangzhou 310015, China}
\affiliation{Department of Physics, Zhejiang University, Hangzhou 310027, China}

\author{Guijun Zhang}
\email{21835019@zju.edu.cn}
\affiliation{School of Mathematics Science, Zhejiang University, Hangzhou 310027, China}

\title{Bures distance of discord for two-qubit X-states}
\begin{abstract}
 Two-qubit X-state is a large class of quantum states which plays an important role in the quantification and dynamical study of quantum correlations. However, the corresponding quantification of quantum discord is still missing for bona fide discord measures, like original quantum discord, Bures distance of discord, and relative entropy of discord. In this paper, we consider the calculation of Bures distance of discord, which is a kind of correlations satisfying all criteria of a discord measure, for two-qubit X-states. Firstly, we derive explicit expression for Bures distance of discord for a kind of five-parameters family of states. Moreover, for general two-qubit X-states, we not only calculate the Bures distance of discord for a subset of two-qubit X-states by classifying and analyzing the optimal local measurements and the optimal projection operators, but also provide an analytic upper bound for entirety.
\end{abstract}
\maketitle

\section{introduction}

Correlations permeate into our interpretation and understanding of the quantum world. Quantum correlations are also regarded as resources needed in quantum algorithms and quantum communication protocols which reveals the quantum advantages over their classical counterpart \cite{Nielsen10,Knill1998,Datta2008A}. To extract correlation information from quantum system, whether classical or quantum, one has to perform measurement. A key difference between the classical and quantum is the characteristics of measurements: while a classical measurement can extract information without disturbance in principle, a quantum measurement often unavoidably breaks the measured system. Actually, quantum measurement lies at the very heart of quantum mechanics, and is the pivotal feature in both theoretical and experimental investigation of quantum information. The theory of non-locality, entanglement, and quantum steering, all depend on quantum measurement \cite{Bruner2014A,HorodeckiRMP09,EPR1935,Gallega2015}.

To some extent, disturbance under quantum measurements signifies quantumness. From the information perspective, the existence of quantum correlation in quantum states will give rise to unavoidable loss of information after quantum measurements. Discord, which was explicitly introduced by Ollivier, Zurek \cite{Zurek2001A} and Henderson, Vedral \cite{Vedral2001A}, to quantify the quantumness of correlations, exactly arises from the loss of information caused by local measurements. The original discord is generalized to multipartite system \cite{Radhakrishnan2020}.

The total correlations (quantum and classical) in  a bipartite quantum system are measured by the quantum mutual information defined as
\begin{align*}
\mathrm{I}(\rho_{AB})=S(\rho_A)+S(\rho_B)-S(\rho_{AB}),
\end{align*}
where $\rho_{A(B)}$ and $\rho$ are the reduced density matrix of subsystem $A(B)$ and the density matrix of the total system, respectively, and $S(\rho):=-\mathrm{Tr}(\rho \mathrm{log}\rho)$ is the von Neumann entropy.

Motivated by the idea that classical correlations are those that can be extracted via quantum measurement, i.e., the maximum amount of correlations extractable by local measurements, a quantification of classical correlations in a bipartite quantum system maybe defined as
\begin{align}\label{eq1}
\mathrm{C}^A(\rho_{AB}):=S(\rho_B)-\mathop{\mathrm{min}}_{\{\pi_i^A\}}\{\sum_ip_iS(\rho_{B|i})\},
\end{align}
where the minimum is over all von Neumann measurements on subsystem A and $p_i=\mathrm{Tr}(\pi_i^A\otimes I)\rho$ is the probability of the measurement outcomes $i$, $\rho_{B|i}=p_i^{-1}\mathrm{Tr}_A(\pi_i^A\otimes I)\rho$ is the corresponding conditional state of B.  The second term in the right of Eq.(\ref{eq1}) represents the minimal remain information of system B after a measurement is made on party A. In other words, $\mathrm{C}^A$ quantifies the maximal amount of information that can be extracted by local measurement on subsystem A. For the direct definition of quantum conditional entropy, i.e., $S(\rho_{AB})-S(\rho_B)$  which can be negative, the left term of Eq.(\ref{eq1}) can be viewed as the corresponding quantum mutual information.

Based on the idea that the total correlations on a quantum system including classical and quantum correlations, therefore, as a measure of quantum correlation, discord can be defined as
\begin{align}\label{eq2}
\delta_A(\rho):=\mathrm{I}(\rho_{AB})-\mathrm{C}^A(\rho_{AB}).
\end{align}
The right part of Eq.(\ref{eq2}) characterizes the amount of mutual information which is not accessible by local measurement on the subsystem A. It can be shown that $\delta(\rho)\ge0$ and $\delta_A(\sigma_{A-cl})=0$ iff
\begin{align*}
\sigma_{A-cl}=\sum_{i=1}^{n_A}p_i\ket{\alpha_i}\bra{\alpha_i}\otimes\sigma_{B|i},
\end{align*}
where $\{\ket{\alpha_i}_i^{n_A}\}$ is an orthonormal basis for subsystem A, $\sigma_{B|i}$ are arbitrary states of subsystem B, and $\{p_i\}$ is probability. We call the whole states A-classical states, and denoted it by $C_A$. To some extend, A-classical states can be viewed as a kind of classical states whose classical correlations can be extracted after a quantum measurements made on subsystem A without any disturbance to the states themselves.

With the set of A-classical states, it is natural to characterize the quantum correlation from a geometric point of view with Bures distance \cite{Bures1969,Uhlmann1976,Uhlmann1986,Uhlmann1995}
\begin{align*}
d_B(\rho,\sigma)=\sqrt{2(1-\sqrt{F(\rho,\sigma)})},
\end{align*}
where $F(\rho,\sigma)=(\mathrm{tr}\sqrt{\sqrt{\rho}\sigma\sqrt{\rho}})^2$ is fidelity \cite{Uhlmann1976}. More than monotonous and Riemannian \cite{Petz1996}, the Bures distance is also jointly convex in state space \cite{Spehner2014}.

The Bures distance of discord is defined as the square of minimal distance to the set of A-classical states \cite{spehner2013A},
\begin{align*}
D_A(\rho):=\min_{\sigma\in C_A}d^2_B(\rho,\sigma).
\end{align*}

It has been proved that $D_A$ is faithful A-classical states, local unitary invariant, non-increasing under local operation in subsystem B, and reduces to an entanglement monotone on pure states \cite{spehner2013A}. And, the evaluation of $D_A$ for the mixed states $\rho$ turns out to be related to an ambiguous quantum state discrimination (QSD) task \cite{Bergou2010,spehner2013A}. Actually, the fidelity between $\rho$ and the closest A-classical state (CCS) is given by the maximum success probability
\begin{align*}
F_A(\rho):=\max_{\sigma\in C_A}F(\rho,\sigma)=\mathop{\mathrm{max}}_{\{\ket{\alpha_i}\}}P^{opt\, v.N}_S(\{\rho_i,\lambda_i\}),
\end{align*}
where $\lambda_i=\bra{\alpha_i}\rho_A\ket{\alpha_i}$ and $\rho_i=\lambda^{-1}_i\sqrt{\rho}\ket{\alpha_i}\bra{\alpha_i}\otimes I\sqrt{\rho}$.

Moreover, denote $\{\ket{\alpha^{opt}_i}\}$ and $\{\ket{\Pi^{opt}_i}\}$ as the basis and projective measurement maximizing $P^{opt v.N}_S(\{\rho_i,\lambda_i\})$, then the CCS(s) to $\rho$ is (are) of the form \cite{spehner2013A}
\begin{align}\label{eq6}
\sigma_{\rho}=\frac{1}{F_A(\rho)}\sum^{n_A}_{i=1}\ket{\alpha^{opt}_i}\bra{\alpha^{opt}_i}\otimes\bra{\alpha^{opt}_i}\sqrt{\rho}\Pi^{opt}_i\sqrt{\rho}\ket{\alpha^{opt}_i}.
\end{align}

Thanks to the link between Bures distance of discord and QSD, for a (2,$n_B$) quantum state $\rho$, one has \cite{spehner2013B}
\begin{align}\label{fidelity-of-states}
F_A(\rho)=\frac{1}{2}\max_{||{\bf u}||=1}\{1-\mathrm{tr}\Lambda({\bf u})+2\sum^{n_B}_{l=1}\lambda_l({\bf u})\}
\end{align}
where $\Lambda({\bf u})=\sqrt{\rho}\sigma_{\bf u}\otimes I\sqrt{\rho}$, $\lambda_l({\bf u})$s are the corresponding eigenvalues in non-increasing order, and $\sigma_{\bf u}:=\sum_{i=1}^3u_m\sigma_m$ for some unit vector ${\bf u}\in R^3$.

Two-qubit X-states, a class of states with natural symmetry structure \cite{Rau2009}, play an important role in studying dissipative dynamical evolution of a quantum system, such as the sudden transitions discussed in \cite{Mazzola2010} and the frozen phenomenon of quantum correlations \cite{Maziero2009,Cianciaruso2015,Adesso2015A}. This class of states includes Werner states \cite{Werner1989} and Bell-diagonal states which also play a key role in entanglement theory. In \cite{luo2008A}, the author calculates the original quantum discord (\ref{eq2}) for Bell-diagonal states. For a general two-qubit X-state, Mazhar Ali \cite{M.Ali2010A} provided an explicit expression for original quantum discord (\ref{eq2}) but later found to be not always correct \cite{huang2013a}. Furthermore, the quantification of Bures distance of discord is still missing with only partial results available for subsets of three parameters \cite{spehner2013B,Adesso2013}. We derive a analytic expression of Bures distance of discord for a large subset of X-states and a tight upper bound is given for the whole class with Eq.(\ref{fidelity-of-states}).

The paper is organised as follows. In section \uppercase\expandafter{\romannumeral2} we compute the Bures distance of discord for two-qubit X-state with $a=d,b=c$ and determine the closest A-classical states of this kind of state. Moreover, we also calculate the Bures distance of classical correlations and find the corresponding closest A-classical state. In section \uppercase\expandafter{\romannumeral3} we discuss closest A-classical state of two-qubit X-states, the optimal measurement and projections of corresponding quantum state discrimination task, and then give the analytic expression of Bures distance of discord for some subsets of two-qubit X-states. We conclude in Section \uppercase\expandafter{\romannumeral4} with a summary and outlook.

\section{Five-parameters X-states}

\subsection{X-states with a=d, b=c}

In this section, let us consider a class of five-parameter family states, two-qubit X-state with $a=d,b=c$. The matrices $\rho$ is given in the standard basis $\{\ket{00},\ket{01},\ket{10},\ket{11}\}$ by
\begin{align}\label{eq13}
\rho=\begin{pmatrix}a&0&0&y\\0&b&x&0\\0&\overline{x}&b&0\\\overline{y}&0&0&a
\end{pmatrix},
\end{align}
with eigenvalues $p_{1(2)}=b\mp|x|,p_{3(4)}=a\mp|y|$ and corresponding eigenvectors
\begin{align*}
&\ket{\phi_{1(2)}}=\frac{1}{\sqrt{2}}(0,\mp1,\overline{x}/|x|,0)^{T},\\
&\ket{\phi_{3(4)}}=\frac{1}{\sqrt{2}}(\mp1,0,0,\overline{y}/|y|)^{T}.
\end{align*}

As $\Lambda({\bf u})$ and $\sigma_{\bf u}\otimes I\rho$ has the same eigenvalues, we can just pay attention to the latter which is easier to calculate. Let ${\bf u}=(\sin\theta\cos\psi,\sin\theta\sin\psi,\cos\theta)$, then in the standard basis, the matrix
\begin{align*}
\sigma_{\bf u}\otimes I\rho=\begin{pmatrix}
am&\overline{nx}&b\overline{n}&my\\
\overline{ny}&bm&mx&a\overline{n}\\
an&m\overline{x}&bm&ny\\
m\overline{y}&bn&nx&am
\end{pmatrix}.
\end{align*}
with $m=\cos\theta,n=e^{i\psi}\sin\theta$ and eigenvalues come in opposite pairs $(\lambda_{\pm}({\bf u}),-\lambda_{\pm}({\bf u}))$,
\begin{align*}
\lambda_{\pm}({\bf u})=\sqrt{\frac{1}{2}\mu\pm\frac{1}{2}\sqrt{\mu^2-4(a^2-|y|^2)(b^2-|x|^2)}},
\end{align*}
where $\mu=\cos^2\theta(a^2+b^2-|x|^2-|y|^2)+2ab\sin^2\theta+2|xy|\sin^2\theta\cos(2\psi+\eta+\xi)$.
Moreover, the fidelity between $\rho$ and $\sigma_{\rho}$ is
\begin{align*}
F_A(\rho)=\frac{1}{2}+\mathop{\mathrm{max}}_{\theta,\psi}(\lambda_+({\bf u})+\lambda_-({\bf u})).
\end{align*}
We notice that if $\mu$ reaches the maximum then it is also true for $F_A(\rho)$. Actually, it is easy to see that
\begin{align*}
(\lambda_+({\bf u})+\lambda_-({\bf u}))^2=\mu+(a^2-|y|^2)(b^2-|x|^2).
\end{align*}
Denoting
$xy=|xy|e^{i\phi},\phi=\eta+\xi$ with $\phi\in[0,2\pi)$,
\begin{align*}
\mu=&\cos^2\theta(a^2+b^2-|x|^2-|y|^2)\\
&+2\sin^2\theta(ab+|xy|\cos(2\psi+\phi))\\
&\leq\cos^2\theta(a^2+b^2-|x|^2-|y|^2)+2\sin^2\theta(ab+|xy|)\\
&=\cos^2\theta((a-b)^2-(|x|+|y|)^2)+2ab+2|xy|
\end{align*}

To maximize $\mu$, one should put $\cos(2\psi+\phi)=1$, i.e., $\psi=-\frac{\phi}{2}$.
As a result,

(\romannumeral1). If $|a-b|>|x|+|y|$, $\mu$ reaches the maximum iff $\cos\theta=1$ which means that
\begin{align}
F_A(\rho)&=\frac{1}{2}+\sqrt{a^2-|y|^2}+\sqrt{b^2-|x|^2}\nonumber\\
\label{eq21}
&=\frac{1}{2}+\sqrt{p_1p_2}+\sqrt{p_3p_4},
\end{align}
and the optimal measurement is $\{\ket{0}\bra{0},\ket{1}\bra{1}\}$.

(\romannumeral2). If $|a-b|<|x|+|y|$, and $|xy|\neq0$, $\mu$ reaches the maximum iff $\cos\theta=0$ and $\psi=-\frac{\phi}{2}$ which means that
\begin{align}\label{eq15}
F_A(\rho)&=\frac{1}{2}+\sqrt{(a+|y|)(b+|x|)}+\sqrt{(a-|y|)(b-|x|)}\nonumber\\
&=\frac{1}{2}+\sqrt{p_2p_4}+\sqrt{p_1p_3},
\end{align}
and the optimal measurement is $\{\frac{1}{2}(I\pm(\cos\frac{\phi}{2}\sigma_1-\sin\frac{\phi}{2}\sigma_2))\}$;
in case of $|xy|=0$ which means that $F_A(\rho)$ take the maximum for any $\phi\in[0,2\pi)$. Then, there are infinite CCS for $\rho$ and the optimal  measurement is $\{\frac{1}{2}(I\pm(\cos\frac{\phi}{2}\sigma_1-\sin\frac{\phi}{2}\sigma_2)),\phi\in[0,2\pi)\}$.

(\romannumeral3). If $|a-b|=|x|+|y|$, (\ref{eq21}) is equal to (\ref{eq15}) and the corresponding optimal measurement depend on whether $|xy|=0$ or not.

1. $|xy|\neq0$, $F_A(\rho)$ reaches the maximum for any $\theta\in[0,\pi]$ and there are infinite CCS with corresponding optimal measurements is $\{\frac{1}{2}(I\pm(\sin\theta\cos\frac{\phi}{2}\sigma_1-\sin\theta\sin\frac{\phi}{2}\sigma_2+\cos\theta\sigma_3)),\theta\in[0,\pi]\} $.

2. $|xy|=0$, $F_A(\rho)$ reaches the maximum for any $\theta\in[0,\pi],\phi\in[0,2\pi)$ and there are infinite CCS with corresponding optimal measurements is $\{\frac{1}{2}(I\pm(\sin\theta\cos\frac{\phi}{2}\sigma_1-\sin\theta\sin\frac{\phi}{2}\sigma_2+\cos\theta\sigma_3)),\theta\in[0,\pi],\phi\in[0,2\pi)\} $.

In conclusion, we have the following theorem.

\begin{thm}
The Bures distance of discord $D_A$ for state of the form (\ref{eq13}) is equal to
\begin{align*}
2(1-\sqrt{\frac{1}{2}+\sqrt{a^2-|y|^2}+\sqrt{b^2-|x|^2}}),\nonumber
\end{align*}
if $|a-b|\ge|x|+|y|$, and $D_A$ is equal to
\begin{align*}
2(1-\sqrt{\frac{1}{2}+\sqrt{(a+|y|)(b+|x|)}+\sqrt{(a-|y|)(b-|x|)}}),\nonumber
\end{align*}
 if $|a-b|<|x|+|y|$.
\end{thm}

\subsection{Closest A-classical states}
Let
\begin{align*}
\sigma_1=\begin{pmatrix}0&1\\1&0\end{pmatrix},\quad
\sigma_2=\begin{pmatrix}0&-i\\i&0\end{pmatrix},\quad
\sigma_3=\begin{pmatrix}1&0\\0&-1\end{pmatrix}
\end{align*}
be the Pauli matrices acting on $\mathrm{C^2}$. Because $\{I,\sigma_1,\sigma_2,\sigma_3\}$ constitutes an operator base for the space of all operators on $\mathrm{C^2}$, any two-qubit state can be written as
\begin{align*}
\rho=&\frac{1}{4}(I\otimes I+\sum_{i}c_{i0}\sigma_i\otimes I+I\otimes\sum_{j}c_{0j}\sigma_j\\
&+\sum_{m,n}c_{mn}\sigma_m\otimes\sigma_n)
\end{align*}
Here $I$ is the identity operator on the composite system or on the component systems, depending on the context.

Denoting $x=|x|e^{i\eta},y=|y|e^{i\xi}$, one can rewrite $\rho$ in the Bloch representation
\begin{align*}
\rho=\frac{1}{4}(I\otimes I+\sum^3_{i=1}c_i\sigma_i\otimes\sigma_i+c_{12}\sigma_1\otimes\sigma_2+c_{21}\sigma_2\otimes\sigma_1),
\end{align*}
with $c_{1(2)}=2(|x|\cos\eta\pm|y|\cos\xi)$, $c_{12(21)}=2(\pm|x|\sin\eta-|y|\sin\xi)$ and $c_3=2(a-b)$. Any such state can be written up to a conjugation by a local unitary $U_A\otimes U_B$ as \cite{Horodecki1996A,luo2009A}
\begin{align*}
\rho^{\prime}=\frac{1}{4}(I\otimes I+\sum^3_{i=1}c^{\prime}_i\sigma_i\otimes\sigma_i)
\end{align*}
where $c^{\prime}_1=2||x|-|y||,c^{\prime}_2=2(|x|+|y|)$ and $c^{\prime}_3=2(a-b)$ without changing the discord. In other words, all this kind of state can be seen as a Bell-diagonal (BD) state up to a local unitary transform. Therefore, we can deduce the CCS of state (\ref{eq13}) with the corresponding result of BD states \cite{spehner2013B}.

\begin{lem}\label{thm2}
 If a quantum state $\rho^{\prime}$ and $\rho$ in bipartite system are invariant up to a local unitary transformation, i.e., $\rho^{\prime}=U_A\otimes U_B\rho U^{\dagger}_A\otimes U^{\dagger}_B$ for some $U_A\otimes U_B$, the same is true for their closest A-classical states.
\end{lem}
\begin{proof}
On one hand, assuming $\{\sigma_{\rho^{\prime}}\}$ are the CCS of $\rho^{\prime}$ and $\sigma_{\rho}$ is a CCS of $\rho$, then
\begin{align}
F(\rho,\sigma_{\rho})&=F(\rho^{\prime},U_A\otimes U_B\sigma_{\rho}U^{\dagger}_A\otimes U^{\dagger}_B)\nonumber\\
&\leq F(\rho^{\prime},\sigma_{\rho^{\prime}})\nonumber\\
&=F(\rho,U^{\dagger}_A\otimes U^{\dagger}_B\sigma_{\rho^{\prime}}U_A\otimes U_B)\nonumber
\end{align}
where in the first and last equality we use the invariance of the fidelity under unitary matrices and in the $"\leq"$ we use the definition of CCS. Because $U^{\dagger}_A\otimes U^{\dagger}_B\sigma_{\rho^{\prime}}U_A\otimes U_B$ is a A-classical state, then the above $"\leq"$ become to $"="$ and
each $U^{\dagger}_A\otimes U^{\dagger}_B\sigma_{\rho^{\prime}}U_A\otimes U_B$ is a CCS of $\rho$.

On the other hand, we want to show that each CCS of $\rho$ can be written as $U^{\dagger}_A\otimes U^{\dagger}_B\sigma_{\rho^{\prime}}U_A\otimes U_B$ for a $\sigma_{\rho^{\prime}}\in\{\sigma_{\rho^{\prime}}\}$. To prove this conclusion, supposing $\sigma^{\prime}_{\rho}$ is a CCS of $\rho$ with $U_A\otimes U_B\sigma^{\prime}_{\rho}U^{\dagger}_A\otimes U^{\dagger}_B\notin\{\sigma_{\rho^{\prime}}\}$. Then,
\begin{align}
F(\rho,\sigma^{\prime}_{\rho})&=F(\rho^{\prime},U_A\otimes U_B\sigma^{\prime}_{\rho}U^{\dagger}_A\otimes U^{\dagger}_B)\nonumber\\
&<F(\rho^{\prime},\sigma_{\rho^{\prime}})\nonumber\\
&=F(\rho,U^{\dagger}_A\otimes U^{\dagger}_B\sigma_{\rho^{\prime}}U_A\otimes U_B)\nonumber\\
&=F(\rho,\sigma_{\rho})\nonumber
\end{align}
where in the first two equality we use the unitary-invariance of fidelity and in the last equality we use the result of the first part of the proof. The $"<"$ is based on the assumption that $U_A\otimes U_B\sigma^{\prime}_{\rho}U^{\dagger}_A\otimes U^{\dagger}_B\notin\{\sigma_{\rho^{\prime}}\}$. The inequality $F(\rho,\sigma^{\prime}_{\rho})<F(\rho,\sigma_{\rho})$ contradict with that assumption which indicates that each CCS of $\rho$ can be written as $U^{\dagger}_A\otimes U^{\dagger}_B\sigma_{\rho^{\prime}}U_A\otimes U_B$ for a $\sigma_{\rho^{\prime}}\in\{\sigma_{\rho^{\prime}}\}.$
\end{proof}

Based on Lemma \ref{thm2} and the corresponding result about BD states in \cite{spehner2013B}, we can deduce the formula of CCS for state (\ref{eq13}):
\begin{align*}
&\sigma_{\rho}(r)=\frac{q^{\prime}_{\mathrm{max}}}{2}[\ket{\alpha^{\prime}_0,\beta^{\prime}_0}\bra{\alpha^{\prime}_0,\beta^{\prime}_0}+\ket{\alpha^{\prime}_0,\beta^{\prime}_0}\bra{\alpha^{\prime}_1,\beta^{\prime}_1}]+\frac{1-q^{\prime}_{\mathrm{max}}}{2}\\
&\times[(1+r)\ket{\alpha^{\prime}_0,\beta^{\prime}_1}\bra{\alpha^{\prime}_0,\beta^{\prime}_1}+(1-r)\ket{\alpha^{\prime}_1,\beta^{\prime}_0}\bra{\alpha^{\prime}_1,\beta^{\prime}_0}]
\end{align*}
if $p^{\prime}_0p^{\prime}_{\mathrm{max}}=0$ and $p^{\prime}_m>0,\forall m\neq m_{max}$, and
\begin{align*}
&\sigma_{\rho}(r)=\frac{q^{\prime}_{\mathrm{max}}}{2}[(1+r)\ket{\alpha^{\prime}_0,\beta^{\prime}_0}\bra{\alpha^{\prime}_0,\beta^{\prime}_0}+(1-r)\ket{\alpha^{\prime}_0,\beta^{\prime}_0}\\
&\bra{\alpha^{\prime}_1,\beta^{\prime}_1}]+\frac{1-q^{\prime}_{\mathrm{max}}}{2}[\ket{\alpha^{\prime}_0,\beta^{\prime}_1}\bra{\alpha^{\prime}_0,\beta^{\prime}_1}+\ket{\alpha^{\prime}_1,\beta^{\prime}_0}\bra{\alpha^{\prime}_1,\beta^{\prime}_0}]
\end{align*}
if $p^{\prime}_0p^{\prime}_{\mathrm{max}}>0$ and $p^{\prime}_1p^{\prime}_2p^{\prime}_3=0$. In this equation $r\in[-1,1]$, $\ket{\alpha^{\prime}_{0(1)}}=U^{\dagger}_A\ket{\alpha_{0(1)}}U_A,\ket{\beta^{\prime}_{0(1)}}=U^{\dagger}_B\ket{\beta_{0(1)}}U_B$ with $\{\ket{\alpha_{0(1)}}, \ket{\beta_{0(1)}} \}$ defined in \cite{spehner2013B} and
\begin{align}
&q_m=\frac{1}{2}+\frac{2\sqrt{p^{\prime}_np^{\prime}_k}-2\sqrt{p^{\prime}_0p^{\prime}_m}+c^{\prime}_m}{4\sqrt{p^{\prime}_np^{\prime}_k}+4\sqrt{p^{\prime}_0p^{\prime}_m}+2},\nonumber\\
&p^{\prime}_0=\frac{1}{4}(1-c^{\prime}_1-c^{\prime}_2-c^{\prime}_3),\nonumber\\
&p^{\prime}_i=\frac{1}{4}(1+c^{\prime}_1+c^{\prime}_2+c^{\prime}_3-2c^{\prime}_i),i=1,2,3,\nonumber
\end{align}
where $\{m,n,k\}$ is a permutation of $\{1,2,3\}$.

\subsection{Classical correlation}

It is different from the quantum entropy case which can be defined with the maximal deviation of quantum entropy after a measurement, the classical correlation is not natural for geometric quantum discord with Bures distance. Denote $CCS_{\rho}$ as the set of all the
CCS of $\rho$, the Bures distance of classical correlation was defined as \cite{Adesso2014A}
\begin{align*}
C_B(\rho):=\inf_{\chi_{\rho}\in CCS_{\rho}}\inf_{\pi\in\mathcal{P}}d^2_B(\chi_{\rho},\pi)=\inf_{\chi_{\rho}\in CCS_{\rho}}d^2_B(\chi_{\rho},\pi_{\chi_{\rho}}),
\end{align*}
with $\mathcal{P}$ is the set of product states, and $\pi_{\chi_{\rho}}$ is any of the closest product states to $\chi_{\rho}$. 

As we can see, for X-state $\rho$ with $a=d,b=c$, we can consider the classical corelation(cc) and the corresponding classical correlated state (ccS). Due to  $\rho^{\prime}=U_A\otimes U_B\rho U^{\dagger}_A\otimes U^{\dagger}_B$ is a BD states, one have
\begin{align}
\mathop{\mathrm{max}}_{\pi\in\mathcal{P}}F(\sigma_{\rho},\pi)=\mathop{\mathrm{max}}_{\pi\in\mathcal{P}}F(\sigma_{\rho^{\prime}},\pi),\nonumber
\end{align}
where in equality we use the fact that $U_A\otimes U_B\pi U^{\dagger}_A\otimes U^{\dagger}_B$ is still a product state for any product state $\pi$.  Obviously, the classical correlated state of this five-parameter family is the same as BD states up to a local unitary. Based on the corresponding result about BD states in \cite{Adesso2014A}, the closest product state $\pi_{\chi_{\rho}}$ for such state (\ref{eq13}) is also $\frac{1}{4}I\otimes I$ and
\begin{align*}
C_B(\rho)=&2-\sum_i\sqrt{p_i}=2-(\sqrt{a+|y|}\\
&+\sqrt{a-|y|}+\sqrt{b+|x|}+\sqrt{b-|x|}).
\end{align*}

\section{General X-states}

For above five-parameters quantum states, we can directly calculate the Bures distance of discord and obtain the closest A-classical state. However, it is not the case for general two-qubit X-states, and we have to discuss the corresponding quantum state discrimination and obtain partial results. In this section, we firstly give the expression of A-classical states and A-classical X-states, and then discuss the optimal measurement and projections of corresponding quantum state discriminations, which results in the analytic equation for some subsets of two-qubit X-states. In the end, we also establish an analytic upper bound.

\subsection{Corresponding A-classical state}

In this section, we will talk about the general formula of A-classical states of two-qubit.
Based on the Bloch representation, each two-qubit state has a one-to-one correspondence to a 15-dimensional vector
\begin{align*}
(c_{10},c_{20},c_{30},c_{01},c_{02},c_{03},c_{11},c_{12},c_{13},c_{21},c_{22},c_{23},c_{31},c_{32},c_{33}),
\end{align*}
sometimes we identify the two-qubit state and corresponding vector. In particular, for a two-qubit X-state, the corresponding vector is
\begin{align*}
(0,0,c_{30},0,0,c_{03},c_{11},c_{12},0,c_{21},c_{22},0,0,0,c_{33}).
\end{align*}

In general, each two-qubit A-classical state can be written as
\begin{align}
\sigma_{A-cl}=p\ket{\alpha_0}\bra{\alpha_0}\otimes\rho_0+(1-p)\ket{\alpha_1}\bra{\alpha_1}\otimes\rho_1,\nonumber
\end{align}
with $p\in[0,1/2]$. Assuming $\ket{\alpha_{0(1)}}\bra{\alpha_{0(1)}}=\frac{1}{2}(\mathrm{I}\pm{\bf r}\cdot\vec{\sigma})$ with ${\bf r}=(r_1,r_2,r_3)$ a unit 3-dimensional real vector  and $\rho_0=\frac{1}{2}(\mathrm{I}+{\bf s}\cdot\vec{\sigma})$, $\rho_1=\frac{1}{2}(\mathrm{I}+{\bf t}\cdot\vec{\sigma})$, any two-qubit A-classical state $\sigma_{A-cl}$ has the following form
 \begin{align}
&\frac{1}{4}[p(\mathrm{I}+{\bf r}\cdot\vec{\sigma})\otimes(\mathrm{I}+{\bf s}\cdot\vec{\sigma})+(1-p)(\mathrm{I}-{\bf r}\cdot\vec{\sigma})\otimes(\mathrm{I}+{\bf t}\cdot\vec{\sigma})]\nonumber\\
&=\frac{1}{4}[\mathrm{I}\otimes\mathrm{I}+\vec{\alpha}\cdot\vec{\sigma}\otimes\mathrm{I}+\mathrm{I}\otimes\vec{\beta}\cdot\vec{\sigma}+{\bf r}\cdot\vec{\sigma}\otimes\vec{\gamma}\vec{\sigma}]\nonumber
\end{align}
with $\vec{\alpha}=(2p-1){\bf r}$, $\vec{\beta}=(p{\bf s}+(1-p){\bf t})$, $\vec{\gamma}=(p{\bf s}-(1-p){\bf t})$. Based on above analysis, we have  the following result.
\begin{lem}\label{thm1}
Each A-classical states of two qubit has following form
\begin{align*}
\frac{1}{4}(I\otimes I+\sum_ic_{i0}\sigma_i\otimes I+\sum_jc_{0j}I\otimes\sigma_j+\sum_{mn}c_{mn}\sigma_m\otimes\sigma_n),
\end{align*}
with
\begin{align}
&c_{i0}=(2p-1)r_i,i=1,2,3, \nonumber \\
&c_{0j}=ps_j+(1-p)t_j,j=1,2,3, \nonumber\\
&c_{mn}=r_m(ps_n-(1-p)t_n),m,n=1,2,3,\nonumber
\end{align}
where $p\in[0,1/2],s_i,r_i,t_i\in\mathit{R}$ and $|{\bf s}|\leq1,|{\bf t}|\leq1,|{\bf r}|=1$.
\end{lem}

\subsection{The form of A-classical X-state}

In this section, we limit our discussion to initially prepared arbitrary two-qubit X-states. The density matrix of a two-qubit X-state in the standard basis $\{\ket{00},\ket{01},\ket{10},\ket{11}\}$ is of the general form
\begin{align*}
\rho_X=\begin{pmatrix}a&0&0&y\\0&b&x&0\\0&\overline{x}&c&0\\\overline{y}&0&0&d
\end{pmatrix},
\end{align*}
with eigenvalues
\begin{align*}
&p_{1(2)}=\frac{1}{2}(b+c\mp\sqrt{(b-c)^2+4|x|^2}), \\
&p_{3(4)}=\frac{1}{2}(a+d\mp\sqrt{(a-d)^2+4|y|^2}),
\end{align*}
and the corresponding eigenvectors are
\begin{align*}
&\ket{\phi_{1(2)}}=(0,(b-c)\mp\sqrt{(b-c)^2+4|x|^2},2\overline{x},0)^{T},\\
&\ket{\phi_{3(4)}}=((a-d)\mp\sqrt{(a-d)^2+4|y|^2},0,0,2\overline{y})^{T}.
\end{align*}

This class of states has an underlying symmetry structure and its CCS is also like this for some special cases.
\begin{lem}\label{thm3}
If a two-qubit X-state has only one closest A-classical state, the CCS is also an X-state.
\end{lem}
\begin{proof}
Assuming $\sigma_{\rho_X}$ is a closest A-classical state of a X-state $\rho_X$, which corresponding vector is
\begin{align*}
(c_{10},c_{20},c_{30},c_{01},c_{02},c_{03},c_{11},c_{12},c_{13},c_{21},c_{22},c_{23},c_{31},c_{32},c_{33}).
\end{align*}
Owing to $\sigma_3\otimes\sigma_3\rho\sigma_3\otimes\sigma_3=\rho$, one has
\begin{align}
F(\rho_X,\sigma_{\rho_X})=F(\rho_X,\sigma_3\otimes\sigma_3\sigma_{\rho_X}\sigma_3\otimes\sigma_3)=F(\rho_X,\sigma^{\prime}_{\rho_X})\nonumber
\end{align}
with
\begin{align}
\sigma^{\prime}_{\rho_X}=(&-c_{01},-c_{02},c_{03},-c_{10},-c_{20},c_{30},\nonumber\\
&c_{11},c_{12},-c_{13},c_{21},c_{22},-c_{23},-c_{31},-c_{32},c_{33}).\nonumber
\end{align}
Obviously, $\sigma^{\prime}_{\rho_X}$ is also a CCS of $\rho_X$. For the assumption that $\rho_X$ has only one CCS, namely $\sigma_{\rho_X}=\sigma^{\prime}_{\rho_X}$, therefore
\begin{align}
\label{eq14}
c_{01}=c_{02}=c_{10}=c_{20}=c_{13}=c_{23}=c_{31}=c_{32}=0,
\end{align}
which means that $\sigma^{\prime}_{\rho_X}$ is also a X-state.
\end{proof}

Replacing the result of Lemma \ref{thm1} into Eq.(\ref{eq14}), one finds that for these A-classical state as the unique CCS for some $\rho_X$,
\begin{align}
&(2p-1)r_1=(2p-1)r_2=0,\nonumber\\
&ps_1+(1-p)t_1=ps_2+(1-p)t_2=0,\nonumber\\
&r_3(ps_1-(1-p)t_1)=r_3(ps_2-(1-p)t_2)=0,\nonumber\\
&r_1(ps_3-(1-p)t_3)=r_2(ps_3-(1-p)t_3)=0.\nonumber
\end{align}
To determine the parameters $c_{ij}$ of the unique $\sigma_{\rho_X}$ for these $\rho_X$, we discuss for different $p$.

(\romannumeral1). For case $0<p<1/2$, one have that $r_1=r_2=0,r_3=1$, and $s_i=\frac{1-p}{p}t_i,i=1,2$. Then,
\begin{align}
&c_{11}=c_{12}=c_{21}=c_{22}=0,c_{33}=ps_3-(1-p)t_3,\nonumber \\
&c_{30}=2p-1,c_{03}=ps_3+(1-p)t_3,\nonumber
\end{align}
and the corresponding A-classical states can be written as
\begin{align*}
\frac{1}{4}\begin{pmatrix} \rho_{11}&0&0&0\\0&\rho_{22}&0&0\\0&0&\rho_{33}&0\\0&0&0&\rho_{44}
\end{pmatrix}
\end{align*}
with $\rho_{11}=1+c_{33}+c_{30}+c_{03},\rho_{22}=1-c_{33}+c_{30}-c_{03},\rho_{33}=1-c_{33}-c_{30}+c_{03},\rho_{44}=1+c_{33}-c_{30}-c_{03}$.

(\romannumeral2). For case $p=0$, it is easy to deduce that $r_1=r_2=t_1=t_2=0,r_3=1$, and then
\begin{align*}
&c_{11}=c_{12}=c_{21}=c_{22}=0,c_{33}=-t_3, \\
&c_{30}=-1,c_{03}=t_3,
\end{align*}
and the corresponding A-classical states is also a diagonal state
with $\rho_{11}=1+c_{33}+c_{30}+c_{03},\rho_{22}=1-c_{33}+c_{30}-c_{03},\rho_{33}=1-c_{33}-c_{30}+c_{03},\rho_{44}=1+c_{33}-c_{30}-c_{03}$.\\

(\romannumeral3). For case $p=1/2$, there has three different case as follows.

1. If $s_3\neq t_3$, one infer that $r_1=r_2=0,r_3=1$ which is also corresponding to the diagonal states.

2. If $s_3=t_3$ and $r_3\neq0$, then $s_1=s_2=t_1=t_2=0$ which means that
\begin{align*}
&c_{03}=s_3,c_{30}=0,\\
&c_{11}=c_{12}=c_{21}=c_{22}=0,c_{33}=0,
\end{align*}
and the corresponding A-classical state is still a diagonal state.\\

3. If $s_3=t_3$ and $r_3=0$, then $s_1=-t_1,s_2=-t_2$ which means that
\begin{align*}
&c_{03}=s_3,c_{30}=0,\\
&c_{11}=r_1s_1,c_{12}=r_1s_2,c_{21}=r_2s_1,c_{22}=r_2s_2,c_{33}=0,
\end{align*}
and the corresponding A-classical states can be written as
\begin{align*}
\frac{1}{4}\begin{pmatrix} \rho_{11}&0&0&\rho_{14}\\0&\rho_{22}&\rho_{23}&0\\0&\rho_{32}&\rho_{33}&0\\\rho_{41}&0&0&\rho_{44}
\end{pmatrix}
\end{align*}
with
\begin{align*}
&\rho_{11}=\rho_{33}=1+c_{03},\rho_{22}=\rho_{44}=1-c_{03}, \\
&\rho_{14}=c_{11}-c_{22}-i(c_{12}+c_{21}),\rho_{41}=c_{11}-c_{22}+i(c_{12}+c_{21}),\\
&\rho_{23}=c_{11}+c_{22}+i(c_{12}-c_{21}),\rho_{32}=c_{11}+c_{22}-i(c_{12}-c_{21}).
\end{align*}

In conclusion, the closest A-classical state for these X-states with only one CCS is neither diagonal state nor X-state. This result will help to derive the corresponding Bures distance of discord for these classes of states with Eq.(\ref{eq6}).

Now, we consider these X-states which have more than one closest A-classical state. In the Section \uppercase\expandafter{\romannumeral2}, the optimal local measurement in subsystem A of CCS has three kind of formula, i.e.,${\bf r}=(\sin\theta\cos\psi,\sin\theta\sin\psi,\cos\theta)$ for\\
1. $\psi\in[0,2\pi)$ with fixed $\theta$,\\
2. $\theta\in[0,\pi],$ with fixed $\psi$, \\
3. $\psi\in[0,2\pi),\theta\in[0,\pi]$.\\

 For X-states $\rho_X$ with $a=d,b=c$, the above $2$ and $3$ is the case. For case 1, if $\theta\neq\{0,\frac{\pi}{2}\}$, maybe no CCS is X-state for these states and we will discuss this situation in next subsection.

 As we can see in Eq.(\ref{eq6}), the CCS of a quantum state $\rho$ depend on both the choice of optimal basis $\ket{\alpha_i^{opt}}$ and optimal projector $\Pi^{opt}$ of $\Lambda(\bf u)$. In fact, for these states which has unique optimal measurement and two different optimal projector, namely $\sum_{i}\ket{\alpha_i}\bra{\alpha_i}\otimes\sigma_{B|i}$ and $\sum_{i}\ket{\alpha_i}\bra{\alpha_i}\otimes\sigma^{\prime}_{B|i}$ are the corresponding CCS, then $p\sum_{i}\ket{\alpha_i}\bra{\alpha_i}\otimes\sigma_{B|i}+(1-p)\sum_{i}\ket{\alpha_i}\bra{\alpha_i}\otimes\sigma^{\prime}_{B|i}$ is also a CCS for $\rho$ for any $p\in[0,1]$\cite{spehner2013B}. Next, we will consider the Bures distance of discord for unique optimal measurement firstly and then pay attention to the different situation of optimal projector.

\subsection{Bures distance of discord for X-states}

Comparing to Eq.(\ref{eq6}), the measurement vector ${\bf u}$ of the optimal measurement is $(0,0,1)$ or $(\cos\psi,\sin\psi,0)$ for a fixed $\psi$ when the CCS of $\rho$ is diagonal states or general X-state. Now, let us estimate the Bures geometric quantum discord and determine the corresponding closest A-classical state.

(\romannumeral1). If the optimal measurement $\ket{\alpha_{0(1)}}\bra{\alpha_{0(1)}}=\frac{1}{2}(I\pm\sigma_3)$, i.e., $\ket{\alpha_{0(1)}}=\ket{0(1)}$. Then, for the eigenvalues and eigenvectors of
\begin{align*}
\sigma_3\otimes I\rho_X=\begin{pmatrix}a&0&0&y\\0&b&x&0\\0&-\overline{x}&-c&0\\-\overline{y}&0&0&-d
\end{pmatrix}
\end{align*}
are $\lambda_{1(2)}=\frac{1}{2}(b-c\mp\sqrt{(b+c)^2-4|x|^2})$, $\lambda_{3(4)}=\frac{1}{2}(a-d\mp\sqrt{(a+d)^2-4|y|^2})$ and
 \begin{align*}
&\ket{\psi_{1(2)}}=(0,(b+c)\mp\sqrt{(b+c)^2-4|x|^2},-2\overline{x},0)^{T},\\
&\ket{\psi_{3(4)}}=((a+d)\mp\sqrt{(a+d)^2-4|y|^2},0,0,-2\overline{y})^{T},
\end{align*}
the fidelity between $\rho_X$ and the CCS is
\begin{align*}
&F^{\prime}_A(\rho_X)=\frac{1}{2}(1+\mathrm{tr}|\Lambda_{\bf u}|)\nonumber \\
&=\frac{1}{2}(1+\sqrt{(b+c)^2-4|x|^2}+\sqrt{(a+d)^2-4|y|^2}).
\end{align*}
We normalize the eigenvectors $\ket{\psi_i}$ and still denote it $\ket{\psi_i}$ for $i=1,2,3,4$.  Due to the $\Lambda({\bf u})=\sqrt{\rho_X}\sigma_3\otimes I\sqrt{\rho_X}$ have the same eigenvalues as $\sigma_3\otimes I\rho_X$ and the corresponding eigenvectors are $\{\sqrt{\rho_X}\ket{\psi_i},i=1,2,3,4\}$, one gets the optimal projector, $\Pi^{opt}=$
\begin{align}\label{eq10}
\left\{
\begin{aligned}
&\sqrt{\rho_X}(\ket{\psi_2}\bra{\psi_2}+\ket{\psi_4}\bra{\psi_4})\sqrt{\rho_X} & \mathrm{if} &\quad bc\neq|x|^2,ad\neq|y|^2\\
&\sqrt{\rho_X}(\ket{\psi_2}\bra{\psi_2}+\ket{\Phi}\bra{\Phi})\sqrt{\rho_X} & \mathrm{if} &\quad bc\neq|x|^2,ad=|y|^2\\
&\sqrt{\rho_X}(\ket{\psi_4}\bra{\psi_4}+\ket{\Psi}\bra{\Psi})\sqrt{\rho_X} & \mathrm{if} &\quad bc=|x|^2,ad\neq|y|^2\\
&\sqrt{\rho_X}(\ket{\Upsilon_1}\bra{\Upsilon_1}+\ket{\Upsilon_2}\bra{\Upsilon_2})\sqrt{\rho_X} & \mathrm{if} &\quad bc=|x|^2,ad=|y|^2
\end{aligned}
\right.
\end{align}
with $\ket{\Phi}\in\mathrm{span}\{\ket{\psi_3},\ket{\psi_4}\}$, $\ket{\Psi}\in \mathrm{span}\{\ket{\psi_1},\ket{\psi_2}\}$ and $\ket{\Upsilon_1},\ket{\Upsilon_2}\in \mathrm{span}\{\ket{\psi_i},i=1,2,3,4\}$, $||\Phi||=||\Psi||=||\Upsilon_1||=||\Upsilon_2||=1$.

If $\det(\rho_X)=(bc-|x|^2)(ad-|y|^2)\neq0$, denoting $\ket{\psi_i}=((\psi_i)_1,(\psi_i)_2,(\psi_i)_3,(\psi_i)_4)^T$, the corresponding closest A-classical state
\begin{align}\label{eq19}
&\sigma_{\rho_X}=\sum^1_{i=0}\ket{i}\bra{i}\otimes\bra{i}\rho_X(\ket{\psi_2}\bra{\psi_2}+\ket{\psi_4}\bra{\psi_4})\rho_X\ket{i}\nonumber\\
&=|\overline{x}(\psi_2)_2+c(\psi_2)_3|^2\ket{00}\bra{00}+|b\overline{(\psi_2)_2}+x\overline{(\psi_2)_3}|^2\ket{01}\bra{01}\nonumber\\
&+|a\overline{(\psi_4)_1}+y\overline{(\psi_4)_4}|^2\ket{10}\bra{10}+|\overline{y}(\psi_4)_1+d(\psi_4)_4|^2\ket{11}\bra{11},
\end{align}
is a diagonal state.

Moreover, even if $\det(\rho_X)=0$, $\sqrt{\rho_X}(\ket{\psi_2}\bra{\psi_2}+\ket{\psi_4}\bra{\psi_4})\sqrt{\rho_X}$ is also a optimal projector and the corresponding CCS is a diagonal state, of course X-state. In other words, supposing the optimal measurement of the CCS for a X-state $\rho_X$ is $\{\ket{0},\ket{1}\}$, we can always find a CCS to be a diagonal state and the corresponding Bures geometric quantum discord
\begin{align*}
D_A(\rho_X)=1-\sqrt{(b+c)^2-4|x|^2}+\sqrt{(a+d)^2-4|y|^2}.
\end{align*}

(\romannumeral2). If the optimal measurement $\ket{\alpha_{0(1)}}\bra{\alpha_{0(1)}}=\frac{1}{2}(I\pm(r_1\sigma_1+r_2\sigma_2))$. Then, on account of the eigenvalues of
\begin{align*}
(r_1\sigma_1+r_2\sigma_2)\otimes I\rho_X=\begin{pmatrix}0&\overline{nx}&\overline{n}c&0\\\overline{ny}&0&0&\overline{n}d\\na&0&0&ny\\0&nb&nx&0
\end{pmatrix}
\end{align*}
are
\begin{align*}
\lambda_{1(2)}=\pm\frac{1}{\sqrt{2}}\sqrt{h+\sqrt{h^2-4(ad-|y|^2)(bc-|x|^2)}} \\
\lambda_{3(4)}=\pm\frac{1}{\sqrt{2}}\sqrt{h-\sqrt{h^2-4(ad-|y|^2)(bc-|x|^2)}}
\end{align*}
where $h=2\mathrm{Re}\{n^2xy\}+ac+bd$ and $n=r_1+ir_2,r_1^2+r_2^2=1$, the fidelity between $\rho_X$ and its CCS can be calculated with Eq.(\ref{fidelity-of-states}):
\begin{align}
F^{\prime\prime}_A(\rho_X)=\max_{{\bf r}=(r_1,r_2,0)}\frac{1}{2}(1+2\lambda_1+2\lambda_3).\nonumber
\end{align}

We notice that if $h$ reaches the maximum then it is also true for $F^{\prime\prime}_A(\rho_X)$. Actually, it is easy to see that
\begin{align}
(\lambda_1+\lambda_3)^2=h+2\sqrt{k},\nonumber
\end{align}
with $k=(ad-|y|^2)(bc-|x|^2)$.

For general $h$, let $r_1=\cos\psi,r_2=\sin\psi,\psi\in[0,2\pi)$ and $xy=|xy|e^{i\phi},\phi\in[0,2\pi]$, then
\begin{align}
h(\psi)&=2|xy|(\cos2\psi\cos\psi-\sin2\psi\sin\phi)+ac+bd,\nonumber\\
&=2|xy|\cos(2\psi+\phi)+ac+bd,\nonumber
\end{align}
and the derivation of $h(\psi)$ is
\begin{align}
\frac{\mathrm{d}h(\psi)}{\mathrm{d}\psi}=-4|xy|\sin(2\psi+\phi).\nonumber
\end{align}

If $|xy|\neq0$, then $h$ reaches the maximum when $\psi=-\frac{\phi}{2}$ , i.e., the vector corresponding to the optimal measurements is $(\cos(\phi/2),-\sin(\phi/2),0)$ with $\phi$ is the phase of $xy$. Therefore, $h_{\mathrm{max}}=2|xy|+ac+bd$ and the fidelity is
\begin{align*}
F^{\prime\prime}_A(\rho_X)=\frac{1}{2}+\sqrt{2|xy|+ac+bd+2\sqrt{(ad-|y|^2)(bc-|x|^2)}}.
\end{align*}

Assuming the eigenvector of $\Lambda({\bf u})$ is $\ket{\psi_i}$ correspond to $\lambda_i$, then the optimal projector can be also represented as (\ref{eq10}). If $bc\neq|x|^2,ad\neq|y|^2$, $\rho_X$ has only one CCS which is an X-state based on the Theorem \ref{thm3}. In the other case, we can also choose $\sqrt{\rho_X}(\ket{\psi_2}\bra{\psi_2}+\ket{\psi_4}\bra{\psi_4})\sqrt{\rho_X}$ as the optimal projector like in case $(0,0,1)$ and the corresponding CCS is an X-state.

If $|xy|=0$, $h$ reaches the maximum for any $\psi\in[0,2\pi]$ which means that there are infinite optimal basis measurement $\frac{1}{2}(I\pm(\cos\psi\sigma_1+\sin\psi\sigma_2))$ for $\psi\in[0,2\pi]$. However,  it is not clear whether there are always exist a X-state CCS for $\rho_X$ in this case.

Therefore, for the X-state $\rho_X$ whose CCS $\sigma_{\rho_X}$ is an X-state, the corresponding fidelity is the maximum of above $F^{\prime}_A(\rho_X)$ and $F^{\prime\prime}_A(\rho_X)$. In other words, denoting $\tau=(b+c)^2-4|x|^2,\kappa=(a+d)^2-4|y|^2$,

(\romannumeral1). if $\sqrt{\tau}+\sqrt{\kappa}\ge2\sqrt{h+2\sqrt{k}}$, i.e., $F^{\prime}_A(\rho_X)>F^{\prime\prime}_A(\rho_X)$, the Bures GQD is
\begin{align}
&D_A(\rho)=2(1-\sqrt{F^{\prime}_A(\rho_X)})\nonumber\\
&=2-\sqrt{2(1+\sqrt{(b+c)^2-4|x|^2}+\sqrt{(a+d)^2-4|y|^2})},\nonumber
\end{align}
and at least one of CCSs is Eq.(\ref{eq19}).

(\romannumeral2). if $\sqrt{\tau}+\sqrt{\kappa}<2\sqrt{h+2\sqrt{k}}$, i.e., $F^{\prime}_A(\rho_X)<F^{\prime\prime}_A(\rho_X)$, the Bures GQD is
\begin{align}
&D_A(\rho_X)=2(1-\sqrt{F^{\prime\prime}_A(\rho_X)})\nonumber\\
&=2-2\sqrt{\frac{1}{2}+\sqrt{2|xy|+ac+bd+2\sqrt{(ad-|y|^2)(bc-|x|^2)}}}.\nonumber
\end{align}

In conclusion, we have the following main theorem.
\begin{thm}
For two qubit X state, if the optimal measurement of the corresponding quantum state discrimination task is $(0,0,1)$ or $(\cos\psi,\sin\psi,0)$ with a unique $\psi$, then there always exist an X-state CCS for $\rho$ and Bures distance of discord is
\begin{align}\label{eq12}
D_A(\rho_X)=\mathrm{max}\{2(1-\sqrt{F^{\prime}_A(\rho_X)}),2(1-\sqrt{F^{\prime\prime}_A(\rho_X)})\},
\end{align}
where
\begin{align*}
&F^{\prime}_A(\rho_X)=\frac{1}{2}(1+\sqrt{(b+c)^2-4|x|^2}+\sqrt{(a+d)^2-4|y|^2}),\\
&F^{\prime\prime}_A(\rho_X)=\frac{1}{2}+\sqrt{2|xy|+ac+bd+2\sqrt{(ad-|y|^2)(bc-|x|^2)}}.
\end{align*}
\end{thm}

In fact, for a general X-state $\rho_X$, it is very difficult to judge whether a measurement is optimal. Therefore, we will try to evaluate Bures distance of discord for X-state by exploring the relationship between the seven parameters in the next subsection.

\subsection{Bures distance of discord based on optimal projector}

This part, we will study the optimal measurement and projector of two-qubit X-state $\rho_X$ with the characteristic polynomial of $\Lambda({\bf u})$. As the $\sigma_{\bf u}\times I\rho_X$ has the same eigenvalues as $\Lambda({\bf u})=\sqrt{\rho_X}\sigma_{\bf u}\times I\sqrt{\rho_X}$, then we focus on the former
\begin{align*}
\sigma_{\bf u}\times I\rho=\begin{pmatrix}ma&\overline{nx}&\overline{n}c&my\\\overline{ny}&mb&mx&\overline{n}d\\na&-m\overline{x}&-mc&ny\\-m\overline{y}&nb&nx&-md
\end{pmatrix},
\end{align*}
where $m=r_3,n=r_1+ir_2$. The corresponding characteristic polynomial of $\Lambda({\bf u})$ is
\begin{align}\label{eq16}
P[\lambda]=\lambda^4+t_3\lambda^3+t_2\lambda^2+ t_1\lambda+t_0,
\end{align}
with
\begin{align}
&t_3=m(-a-b+c+d),\nonumber\\
&t_2=m^2(ab-bc-ad+cd+|x|^2+|y|^2)-h,\nonumber\\
&t_1=m[(a-d)(bc-|x|^2)+(b-c)(ad-|y|^2)],\nonumber\\
&t_0=(ad-|y|^2)(bc-|x|^2),\nonumber
\end{align}
where $h=2\mathrm{Re}\{n^2xy\}-ac-bd$. The coefficient $t_2$ is the only one depend on both $m$ and $n$, and the constant term of $P[\lambda]$ is the determinant of $\rho$, i.e. $t_0=\det(\rho)$. Supposing $\lambda_1\ge\lambda_2\ge\lambda_3\ge\lambda_4$ are the eigenvalues of $\Lambda({\bf u})$, the $\Pi^{opt}$ is the spectral projector associated to the two highest eigenvalues, i.e., $\lambda_1,\lambda_2$, on the basis of the result of QSD \cite{Spehner2014}.

 If $\det(\rho_X)=0$, namely $(1)ad=|y|^2$ or $bc=|x|^2$ or both, then at least one of the eigenvalue is $0$ and the state $\rho_X$ has infinite optimal projectors. Moreover, if $t_1=0$ also holds, i.e. $ad=|y|^2$ and $bc=|x|^2$, $a=d=|y|$ or $b=c=|x|$, two non-zero real roots of $P[\lambda]$ are  $\lambda_{1(4)}=\frac{-t_3\pm\sqrt{t_3^2-4t_2}}{2}$. Therefore, the fidelity between $\rho$ and its CCS $F(\rho_X)=\frac{1}{2}+\mathrm{max}_m\{\lambda_1(m)\}$ with
 \begin{align}
 2\lambda_1(m)=\sqrt{m^2g(a,b,c,d)+8|xy|+4ac+4bd}-m\Delta\nonumber
 \end{align}
 and the derivation of $2\lambda_1(m)$ is
 \begin{align}
 \frac{2d\lambda_1(m)}{dm}=\frac{mg(a,b,c,d)}{\sqrt{m^2g(a,b,c,d)+8|xy|+4ac+4bd}}-\Delta\nonumber
 \end{align}
 where $g(a,b,c,d)=2(a^2+b^2+c^2+d^2)-1-4(|x|^2+|y|^2-ad-bc)-8|xy|$ and $\Delta=c+d-a-b$ .
 Denoting $g=g(a,b,c,d)$ , to get the maximum eigenvalue $\lambda_1$, the corresponding optimal $m$ is
\begin{align}
m_{opt}=\left\{
\begin{aligned}
&1 &\quad g\geq0,\Delta<0\\
&0 &\quad g\leq0,\Delta\geq0\\
&0\&1 &\quad g\geq0,\Delta\geq0\\
&\frac{-2\sqrt{2|xy|+ac+bd}\Delta}{\sqrt{g^2-(c+d-a-b)^2g}} & g<0,\Delta<0\\
\end{aligned}
\right.\nonumber
\end{align}
 Therefore, the corresponding fidelity $F(\rho_X)$ is
\begin{align*}
\left\{
\begin{aligned}
&\frac{1+\sqrt{2(a+c)^2+2(b+d)^2-1}}{2}-\Delta &\quad  g\geq0,\Delta<0\\
&\frac{1}{2}+\sqrt{ac}+\sqrt{bd} &\quad g\leq0,\Delta\geq0\\
&\frac{1}{2}+\mathrm{max}\{\lambda_1(0),\lambda_1(1)\} &\quad g\geq0,\Delta\geq0\\
&\frac{1}{2}+\lambda_1(-\frac{2\sqrt{2|xy|+ac+bd}\Delta}{\sqrt{g^2-(c+d-a-b)^2g}})&\quad g<0,\Delta<0\\
\end{aligned}
\right.
\end{align*}

 If $\det(\rho_X)>0$, then $\lambda_1\ge\lambda_2>0$ and the optimal projector is unique. Therefore, the number of corresponding CCS is just dependent on the optimal measurement.

 (\romannumeral1). In case the CCS is unique, so is the optimal measurement $\ket{\alpha^{opt}_i}$. This situation has been dealt with in Section \uppercase\expandafter{\romannumeral3}.C.

 (\romannumeral2). In case the CCS is infinite, so is  the optimal measurement $\ket{\alpha^{opt}_i}$. and the corresponding measurement should be
 $(\sin\theta\cos\psi,\sin\theta\sin\psi,\cos\theta),\psi\in[0,2\pi)$ for fixed $\theta$. For example, if $|xy|=0$, then the coefficients of  $P[\lambda]$ are all independent of the value of $\psi$ which means that the optimal measurement is happen to be
 $(\sin\theta\cos\psi,\sin\theta\sin\psi,\cos\theta),\psi\in[0,2\pi)$ for some $\theta$.

Furthermore, if $a=d,b=c$, the X-state reduces to (\ref{eq13}). In this case, $t_3=t_1=0$, and then $\lambda_1=\sqrt{-\frac{t_2}{2}+\frac{\sqrt{t_2^2-4t_0}}{2}}$ which consistence to the result in Section \uppercase\expandafter{\romannumeral2}.

\subsection{An analytic upper bound}

For general X-states, it is not the case that there exist an X-state CCS for each $\rho_X$. In fact, since the A-classical X-states is the subset of A-classical states, the minimal Bures distance between $\rho_X$ and the set of A-classical X-state provides an upper bound for Bures distance of discord for X-state $\rho_X$, namely,
\begin{align}
\mathop{\mathrm{min}}_{\sigma_{A-cl}\in C_A}d_B(\rho_X,\sigma_{A-cl})^2\leq\mathop{\mathrm{min}}_{\sigma_{A-cl}\in C_{A_X}}d_B(\rho_X,\sigma_{A-cl})^2\nonumber
\end{align}
where $C_{A_X}$ is the set of all  A-classical X states. Obviously, this inequality becomes to equality for these states which have X-state CCS. In addition, the Bures distance of discord can be indeed an upper bound for some case. For example, an X-state with $a=b=\frac{1}{3}$, $ |x|=|y|=c=d=\frac{1}{6}$, then
\begin{align}
&g=-\frac{4}{9},\Delta=-\frac{1}{3},m_{opt}=\sqrt{\frac{3}{10}},\nonumber\\
&\lambda_1(0)=\frac{1}{\sqrt{6}},\lambda_1(1)=\frac{\sqrt{2}+1}{6},\lambda_1(m_{opt})=\sqrt{\frac{5}{24}}.\nonumber
\end{align}
The fact, $\lambda_1(m_{opt})>\lambda_1(0)>\lambda_1(1)$, indicates that the right side of Eq.(\ref{eq12}) is a strict upper bound for Bures distance of discord for this state. In other words, the optimal measurement for two-qubit X-states is not always $(0,0,1)$ and $(\cos\psi,\sin\psi,0)$.

\section{conclusion}

How can we meaningfully quantify quantum correlations in arbitrary quantum states? This question lies at the very heart of quantum correlation theory. In this paper, we evaluate the Bures distance of discord for two-qubit X-states with its connection to quantum state discriminations \cite{spehner2013A,spehner2013B}. For a subset of two-qubit X-states, we derive the explicit expression for both quantum and classical correlation in the perspective of geometric, and determine the corresponding closest zero-discord states and zero-correlation states (product states).

For general X-states, on one hand, we calculate the Bures distance of discord for these states which has a unique CCS based on the fact that the unique CCS for X-state must be also an X-state. On the other hand, an explicit expression for Bures distance of discord is given for these states whose corresponding characteristic polynomial (\ref{eq16}) has only two non-zero roots. In addition, we provide an upper bound for the Bures distance of discord for general X-states based on the minimal Bures distance between the X-state and the set of X-state closest A-classical states.

This generalized result is previously available only for a three-parameter subset of such states. There we maximize the fidelity with the help of the result from quantum state discrimination, it would be of interest to explore another method to calculate the maximum of fidelity which will be helpful for quantum state discriminations and vice versa.

\begin{acknowledgments}
{\em  This project is supported by the National Natural Science Foundation of China (Grants No. 61877054), and Jiangxi Provincial Natural Science Foundation (20202BAB201001).
}
 \end{acknowledgments}


%

\end{document}